\documentclass[aps,prb,twocolumn,superscriptaddress,showpacs]{revtex4}

\usepackage{graphicx}
\usepackage{color}

\begin{document}

\title{Pressure- and Field-Tuning the Magnetostructural Phases of Mn$_3$O$_4$:  Raman Scattering and X-Ray Diffraction Studies}
\author{M. Kim}
\author{X.M. Chen}
\author{X. Wang}
\affiliation{Department of Physics and Frederick Seitz Materials
Research Laboratory, University of Illinois, Urbana, Illinois 61801}
\author{C.S. Nelson}
\affiliation{National Synchrotron Light Source, Brookhaven National Laboratory
Upton, NY 11973}
\author{R. Budakian}
\author{P. Abbamonte}
\author{S.L. Cooper}
\affiliation{Department of Physics and Frederick Seitz Materials
Research Laboratory, University of Illinois, Urbana, Illinois 61801}

\date{\today}

\begin{abstract}
We present temperature-, magnetic-field-, and pressure-dependent Raman scattering studies of single crystal Mn$_3$O$_4$, combined with temperature- and field-dependent x-ray diffraction studies, revealing the novel magnetostructural phases in Mn$_3$O$_4$. Our temperature-dependent studies showed that the commensurate magnetic transition at $T_2$=33K in the binary spinel Mn$_3$O$_4$ is associated with a structural transition from tetragonal to orthorhombic structures. Field-dependent studies showed that the onset and nature of this structural transition can be controlled with an applied magnetic field, and revealed evidence for a field-tuned quantum phase transition to a tetragonal spin-disordered phase for \textbf{H}$\parallel$[$1\overline{1}0$]. Pressure-dependent Raman measurements showed that the magnetic easy axis direction in Mn$_3$O$_4$ can be controlled---and the ferrimagnetic transition temperature increased---with applied pressure. Finally, combined pressure- and magnetic-field-tuned Raman measurements revealed a rich magnetostructural phase diagram---including a pressure- and field-induced magnetically frustrated tetragonal phase in the \emph{PH} phase diagram---that can be generated in Mn$_3$O$_4$ with applied pressure and magnetic field.
\end{abstract}


\maketitle

\section{Introduction}
Magnetic spinels are well-known to exhibit geometrical frustration, wherein lattice geometry conflicts with the magnetic interactions that favor a particular minimum energy state. Such frustration can create macroscopic ground state degeneracies, the absence of long range order, and the appearance of novel phenomena and states of matter~\cite{ref1,ref2,ref3} that are of both scientific and technological interest: for example,  sulfur-based spinels such as Fe$B_2$S$_4$ (\emph{B}=Cr,Sc)~\cite{ref4,ref5} exhibit novel low-temperature phases such as orbital-liquid or -glass ground states in which frustration prevents long-range orbital order down to \emph{T}=0; the chromium-oxide spinels $AB_2$O$_4$ (\emph{A}=Zn, Cd, Hg; \emph{B}=Cr) exhibit three-dimensional spin-Peierls-like transitions where lattice distortions relieve spin frustration;~\cite{ref6,ref7,ref8,ref9,ref10} and the vanadium-oxide spinels $AB_2$O$_4$ (\emph{A}=Zn, Cd, Mn; \emph{B}=V) display complex spin/orbital ordering which are associated with large magnetoelastic and magnetodielectric effects.~\cite{ref11,ref12,ref13,ref14,ref15,ref16,ref17,ref18} This diversity of low temperature behavior is generally recognized as governed by the delicate interplay between geometrical frustrated exchange interactions, spin-orbital coupling, and applied magnetic field.~\cite{ref19,ref20} However, there has been little experimental investigation of the microscopic details of this interplay; moreover, the role that magnetoelastic interactions play in governing low temperature behavior of spinels has not been adequately explored, particularly given the prevalence of structural transitions and magnetoelastic anomalies in these materials.

The binary spinel Mn$_3$O$_4$ is an ideal model system in which to investigate the complex interplay between structure, competing interactions, spin-orbital-lattice coupling, and applied magnetic field in the spinels; this is because Mn$_3$O$_4$---in spite of its simpler chemical composition with Mn ions at both tetrahedral (\emph{A}=Mn$^{2+}$) and octahedral (\emph{B}=Mn$^{3+}$) sites (see Fig. 1(a))---is known to exhibit the rich low-temperature behaviors characteristic of more complex ternary magnetic spinels. For example, Mn$_3$O$_4$ shows three different magnetic transitions with complex spin ordering at low temperatures: below $T_C$=43K, the spins in Mn$_3$O$_4$ exhibit Yafet-Kittel-type ferrimagnetic ordering, in which the net spin of the octahedrally coordinated Mn$^{3+}$ spins is antiparallel to the [110] direction of the tetrahedrally coordinated Mn$^{2+}$ spins, with pairs of Mn$^{3+}$ spins canted by ${\pm}{\Theta}_{YK}$ from the [$\overline{1}\overline{1}0$] direction (see Fig. 1(b)), where cos${\Theta}_{YK}$=0.38 (0.33) at \emph{T}=4.7K and cos${\Theta}_{YK}$=0.45 (0.20) at \emph{T}=29K for the ``non-doubling'' octahedral (``doubling'' octahedral) site.~\cite{ref18,ref21}  However, below $T_1$=39K, Mn$_3$O$_4$ develops an incommensurate sinusoidal or spiral spin structure of the Mn$^{3+}$ spins; and below $T_2$=33K, Mn$_3$O$_4$ exhibits a commensurate spin structure in which the magnetic unit cell doubles the chemical unit cell.~\cite{ref18,ref21,ref22} Recent studies also show that the magnetic transitions in Mn$_3$O$_4$ are associated with significant changes in the dielectric constant, indicating strong magnetodielectric coupling;~\cite{ref23,ref24} indeed, the strong spin-lattice coupling implied by these results is confirmed by observations that the dielectric constant in Mn$_3$O$_4$ is strongly dependent upon both the magnitude and direction of an applied magnetic field.~\cite{ref24} The dielectric changes in Mn$_3$O$_4$ are also accompanied by strain changes,~\cite{ref24} suggesting that the complex magnetic ordering and magnetodielectric/magnetoelastic behaviors in Mn$_3$O$_4$ are associated with magneto-structural changes induced by magnetic ordering or magnetic field. However, due to a dearth of microscopic measurements, the microscopic details associated with the low-temperature behaviors in Mn$_3$O$_4$---particularly the microscopic mechanism by which spin-phonon coupling mediates the magnetodielectric behavior in Mn$_3$O$_4$---remain uncertain.

\begin{figure}[tb]
 \centerline{\includegraphics[width=8.5cm]{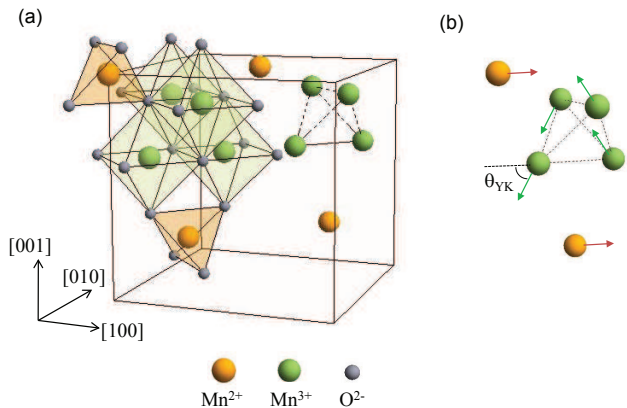}}
 \caption{\label{fig1}(a) Schematic representation of the spinel structure of Mn$_3$O$_4$: Mn$^{2+}$ ions are located in the tetrahedrally coordinated sites; and, Mn$^{3+}$ ions are located in the octahedrally coordinated sites. Also, the Mn$^{3+}$ sites (i.e., octahedral sites) form corner-sharing tetrahedra, resulting in geometric frustration. (b) Schematic representation of the spin structure of Mn$_3$O$_4$ in Yafet-Kittel-type ferrimagnetic ordering: ${\Theta}_{YK}$ is a canting angle of Mn$^{3+}$ spins from the [$\overline{1}\overline{1}0$] direction.}
\end{figure}

To elucidate the microscopic details underlying the temperature- and magnetic-field-dependent phases and properties in Mn$_3$O$_4$, we have used temperature-, magnet-field-, and pressure-dependent inelastic light (Raman) scattering and temperature- and field-dependent x-ray scattering to investigate the temperature-, field-, and pressure-dependent phases of this material. Raman scattering is particularly effective for studying the complex phases of materials such as Mn$_3$O$_4$ because (i) it conveys important information regarding the microscopic structural and magnetic  phases of materials, by providing detailed energy, symmetry, and lifetime information about the electronic, magnetic, and lattice degrees-of-freedom in a material; and it can be readily performed under the extreme conditions needed for exploring low temperature, high magnetic field, and high pressure phase changes.~\cite{ref25,ref26}

In the Raman studies of Mn$_3$O$_4$ presented here, spectroscopic ``maps'' of the phonon frequency ($\omega$) vs. temperature (\emph{T}), applied pressure (\emph{P}), and magnetic field (\emph{H}) are measured in Mn$_3$O$_4$, from which detailed \emph{HT}-, \emph{PT}- and \emph{PH}-diagrams of the magnetostructural phases of Mn$_3$O$_4$ are obtained. These Raman and x-ray studies allow us to elucidate the microscopic structural changes that underlie the low-temperature magnetic ordering and magnetodielectric behavior and to uncover new magnetostructural phase regimes in which magnetodielectric and other novel phenomena are observed. Among the important results of our studies: (i) temperature-dependent studies show that the cell-doubled commensurate magnetic transition at $T_2$=33K in Mn$_3$O$_4$ is associated with a structural transition from tetragonal to orthorhombic structures, which lifts the degeneracy in the \emph{a-b} plane. This result reveals the mechanism by which Mn$_3$O$_4$ relieves geometrical frustration and magnetically orders at low temperatures. (ii) Field-dependent studies show that the onset and nature of this structural transition depends on both the magnitude and the direction of the magnetic field, elucidating the microscopic origins of the remarkable magneto dielectric and  elastic properties of Mn$_3$O$_4$. Of particular interest is the observation of a field-tuned quantum phase transition to a tetragonal spin-disordered phase for magnetic fields applied transverse to the ferrimagnetic easy-axis direction, i.e., for \textbf{H}$\parallel$[$1\overline{1}0$], which we argue arises from a field-tuned degeneracy of magnetostructural states in Mn$_3$O$_4$. (iii) Pressure-dependent Raman measurements show that the magnetic easy axis direction in Mn$_3$O$_4$ can be controlled---and the ferrimagnetic transition temperature increased---with applied pressure. (iv) Finally, combined pressure- and magnetic-field-tuned Raman measurements reveal a rich magnetostructural phase diagram---including a pressure- and field-induced magnetically frustrated tetragonal phase in the \emph{PH} phase diagram---that can be generated in Mn$_3$O$_4$ with applied pressure and magnetic field. Overall, our studies of this relatively simple binary spinel illustrate the importance of magnetoelastic energies---which have been largely ignored in theoretical investigations---in governing the low temperature phase behavior of the spinels.

\begin{figure}[tb]
 \centerline{\includegraphics[width=8.5cm]{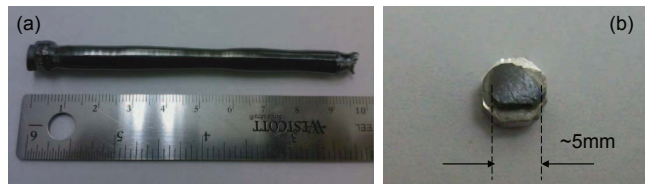}}
 \caption{\label{fig2}(a) A whole volume of the result Mn$_3$O$_4$ crystal grown using the floating zone technique. (b) Sample with surface normal along the [110] direction.}
\end{figure}

\section{Experimental procedure}
\subsection{Sample preparation}
The strong dependence of the dielectric properties and strain on the applied magnetic field direction relative to the crystal axes of Mn$_3$O$_4$---which is related to this material's anisotropic crystal and spin structure---underscores the importance of using a single crystal for studying this material.  Therefore, all our measurements were performed on single crystal Mn$_3$O$_4$. A single-crystal sample of Mn$_3$O$_4$ was grown at the University of Illinois for this study, using an optical floating zone (OFZ) technique as follows:

Commercially available fine Mn$_3$O$_4$ powder (manganese (II, III) oxide, -325 mesh, 97$\%$, Sigma-Aldrich Co.) was used as the starting material.  The powder was formed into a cylindrical shape with a diameter of $\sim$10mm and a length of $\sim$150mm, and was pressed at a hydrostatic pressure of $\sim$40MPa. The pressed rods were sintered at 1050$^{\circ}$C for 5 hours with an argon gas flow of 0.5 L/min. These sintered rods were used as the feed/seed rods---which are a prerequisite for crystal growth using the OFZ method. Single crystal growth was performed using a four-ellipsoid-mirror OFZ furnace (Crystal Systems Inc. FZ-T-10000-H-VI-VP) equipped with four 1000W halogen lamps. The growth conditions used included: a heating power of 58.5$\%$ of the 1000-W lamp; a feed/seed (upper/lower shaft) rotation rate of 35 rpm in opposite directions; a growth rate (mirror-stage moving rate) of 5 mm/h; a feeding rate (upper-shaft moving rate) of 1 mm/h; and a growth atmosphere of oxygen at a gas pressure of 0.1 MPa.

Temperature-, and magnetic-field-dependent Raman and x-ray scattering measurements were performed on the as-grown surfaces of single crystal samples with diameters of 5-10 mm and surface normals along the [110] direction---one of the sample pieces used for the Raman measurements is shown in Fig. 2(b)---which were cut from the full crystalline rod shown in Fig. 2(a).

\begin{figure}[tb]
 \centerline{\includegraphics[width=8.5cm]{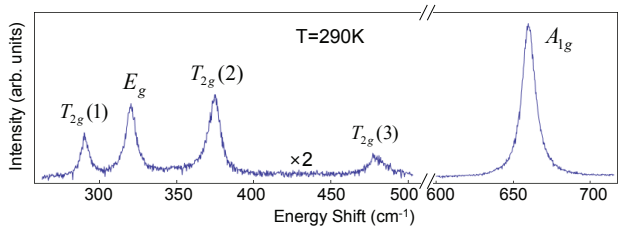}}
 \caption{\label{fig3}Room-temperature Raman spectrum of Mn$_3$O$_4$.}
\end{figure}

\subsection{Sample characterization}
The stoichiometry of the grown crystal was verified with powder XRD measurements, and the resulting spectrum matched well with the reference pattern of Mn$_3$O$_4$. The crystalline orientation of the samples used for measurements was determined from an XRD pole figure analysis using a Phillips X'pert system. The sample quality was further demonstrated by the room temperature Raman spectrum shown in Fig. 3---which clearly exhibits 5 Raman modes at 290 cm$^{-1}$, 320 cm$^{-1}$, 375 cm$^{-1}$, 479 cm$^{-1}$, and 660 cm$^{-1}$---consistent with previous reports.~\cite{ref27,ref28,ref29} Magnetic susceptibility measurements---performed using a Quantum Design SQUID Magnetometer (MPMS-7T XL) at the University of Kentucky---also confirmed the high quality of the samples. As seen in Fig. 4, the results clearly show three different magnetic transitions, consistent with previous measurements:~\cite{ref18,ref21,ref22} these include the magnetic transitions to the Yafet-Kittel ferrimagnetic phase at \emph{T}=43K, to the incommensurate sinusoidal/spiral spin phase at \emph{T}=39K, and to the cell-doubled ferrimagnetic phase at \emph{T}=33K.

\begin{figure}[tb]
 \centerline{\includegraphics[width=8.5cm]{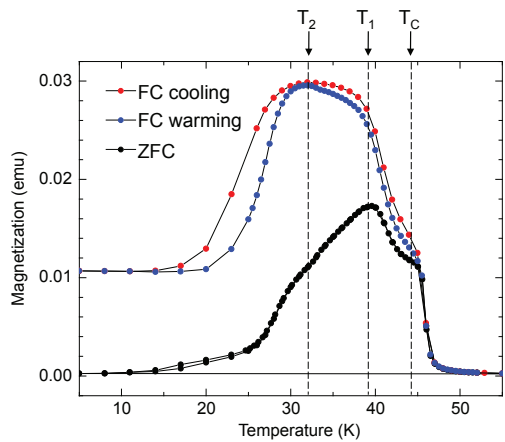}}
 \caption{\label{fig4}Magnetic properties of the Mn$_3$O$_4$ crystal. The temperature dependence of magnetization shows three different magnetic transitions, which are consistent to the previously known ones: a transition at $T_C$=43K from high-temperature paramagnetic phase to Yafet-Kittel ferrimagnetic phase, a transition at $T_1$=39K to an incommensurate phase, and a transition at $T_2$=33K to the cell-doubled commensurate phase.}
\end{figure}

\subsection{Raman measurements}
In this study, Raman scattering was used as the main experimental method to probe the magnetostructural phases of Mn$_3$O$_4$ at low temperatures, high-magnetic fields, and high pressures. The Raman scattering measurements were performed using the 647.1 nm excitation line from a Kr$^+$ laser. The incident laser power was limited to 10mW, and was focused to a ${\sim}$50${\mu}$m-diameter spot to minimize laser heating of the sample. The scattered light from the samples was collected in a backscattering geometry, dispersed through a triple stage spectrometer, and then detected with a liquid-nitrogen-cooled CCD detector. The incident light polarization was selected with a polarization rotator, and the scattered light polarization was analyzed with a linear polarizer, providing symmetry information about the excitations studied. The samples were inserted into a continuous He-flow cryostat, which itself was horizontally mounted in the open bore of a superconducting magnet as described in Fig. 5. This experimental arrangement allowed for Raman scattering measurements under the simultaneous conditions of low temperature (3-290 K), high magnetic field (0-9 T), and high pressure (0-100 kbar).

Magnetic field measurements were performed in both Voigt ($\overrightarrow{q}$$\perp$\textbf{H}) and Faraday ($\overrightarrow{q}$$\parallel$\textbf{H}) geometries, where $\overrightarrow{q}$ is the wavevector of the incident light and \textbf{H} is the magnetic field direction. The field measurements in the Faraday geometry were performed by mounting the samples at the end of the insert as illustrated in Fig. 5(a), so that  the wavevector of the incident light is parallel to the applied field. On the other hand, the Voigt geometry was achieved by mounting the sample in the following way: the sample was mounted on an octagon plate, which is mounted sideways on the sample rod, as illustrated in Fig. 5(b).  The incident light was guided to the same sample surface with a 45 degree mirror mounted on the sample rod near the sample. This sample mounting allows the magnetic field to be applied perpendicular to the wavevector of the incident light.

High-pressure measurements were performed using a MCDAC-type high-pressure cell---in which  moissanite (SiC) anvils were used to provide pressure on the sample via an argon liquid medium---inserted into the cryostat as illustrated in Fig. 5(c). This high pressure cell allows pressure control with the cell inside the cryostat, so that the pressure can be changed \emph{in situ} at low temperatures without any extra warming/cooling procedure. Also, this arrangement allows simultaneous magnetic-field measurements in the Faraday ($\overrightarrow{q}$$\parallel$\textbf{H}) geometry, as illustrated in Fig. 5(c). The pressure was determined at low temperatures and high magnetic fields from the shift in the fluorescence line of a ruby chip loaded in the cell along with the sample piece.

\begin{figure}[tb]
 \centerline{\includegraphics[width=8.5cm]{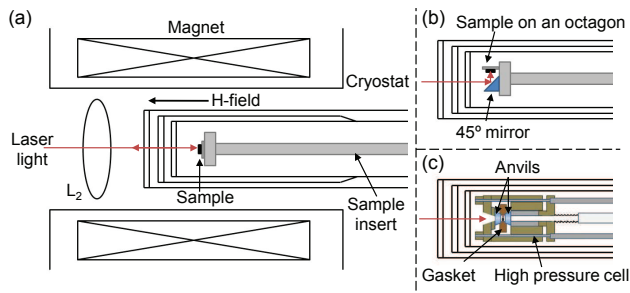}}
 \caption{\label{fig5}Illustrations of the apparatus arrangement for high-magnetic-field and/or high-pressure measurements at low temperatures. (a) For high magnetic field measurements with the Faraday geometry, the sample is mounted on the face of the sample holder using silver paint, facing the optical wind window of the cryostat. (b) For high magnetic field measurements with the Voigt geometry, the sample is mounted in an octagon plate, and it is mounted sideways on the sample rod. A 45-degree mirror is used to guide the incident light to the sample. (c) For high pressure measurements combined with the simultaneous high-magnetic-field measurements, high-pressure cell is inserted into the horizontal cryostat inside the open bore of the superconducting magnet.}
\end{figure}

\subsection{X-ray measurements}
To check the structural information obtained using our Raman results, x-ray scattering measurements were also performed on the same single crystal sample of Mn$_3$O$_4$. Temperature-dependent x-ray measurements were carried out using Mo K$_\alpha$ radiation from a Rigaku rotaflex RU-300 rotating anode source. A graphite (004) monochromator was employed to remove Bremsstrahlung radiation from the source, and an energy-resolving, solid state detector was used with a multichannel analyzer to reject sample fluorescence. A closed cycle He refrigerator was used to control the sample temperature in the range of 10K to 75K. A Phillips MRD X'Pert system was used for high precision measurements at room temperature. A least squares program was used to determine the lattice parameters of the crystal from the data. Magnetic-field-dependent x-ray measurements were also perform with vertical field and horizontal scattering plane---i.e. a scattering geometry of $\overrightarrow{q}$$\perp$\textbf{H}---at Beamline X21, National Synchrotron Light Source (NSLS), Brookhaven National Laboratory, allowing structural measurements of Mn$_3$O$_4$ at low temperatures down to 1.8 K and high magnetic fields up to 10T.

\section{Results and discussion}

\subsection{Temperature dependence}
Fig. 3 shows the room-temperature Raman spectrum of Mn$_3$O$_4$, which exhibits 5 phonon peaks: a (triply degenerate) $T_{2g}$ symmetry mode at 290 cm$^{-1}$, a (doubly degenerate)  $E_g$ symmetry mode at 320 cm$^{-1}$, $T_{2g}$ symmetry modes at 375 cm$^{-1}$ and 479 cm$^{-1}$, and a (singly degenerate) $A_{1g}$ symmetry ``breathing'' mode at 660 cm$^{-1}$.~\cite{ref30}  In this paper, we focus on the lowest energy $T_{2g}$ symmetry mode, designated as $T_{2g}$(1), which is known from previous Raman studies to be associated with Mn-O bond-stretching vibrations of the tetrahedral site ions (Mn$^{2+}$ in Mn$_3$O$_4$);~\cite{ref28,ref29,ref31,ref32} therefore, this phonon mode can provide detailed information about structural changes associated with the tetrahedral site and the Mn$^{2+}$-O$^{2-}$ bond-lengths.

Figs. 6(a) and (c) show the temperature dependence of the $T_{2g}$(1) mode for light polarized along the [$1\overline{1}0$] crystallographic direction of Mn$_3$O$_4$, from which three distinct temperature regimes in Mn$_3$O$_4$ can be identified. Above the transition temperature to the ferrimagnetic phase at $T_C$=43K, the $T_{2g}$(1) mode shifts from $\sim$290 to $\sim$295 cm$^{-1}$ in a conventional manner consistent with anharmonic effects;~\cite{ref33} further, although there is a slight decrease in the $T_{2g}$ mode frequency between the cell-doubling transition temperature at $T_2$=33 K and $T_C$=43K, presumably due to magnetoelastic effects, this mode exhibits no evidence for a change in structural symmetry above $T_2$=33K. On the other hand, below $T_2$=33 K, the $T_{2g}$(1) mode abruptly splits into two modes near 290 cm$^{-1}$ and 300 cm$^{-1}$, where the lower energy ($\sim$290 cm$^{-1}$) split mode reflects an  expansion of the Mn$^{2+}$-O$^{2-}$ bond length and the higher energy ($\sim$300 cm$^{-1}$) split mode reflects a contraction of the Mn$^{2+}$-O$^{2-}$ bond length.~\cite{ref32} Furthermore, the 300 cm$^{-1}$ split mode shows a much stronger intensity while the 290 cm$^{-1}$ split mode has a substantially weaker intensity, indicating that the vibrations of the contracted Mn$^{2+}$-O$^{2-}$ bond are more strongly coupled to the incident light polarized along the [$1\overline{1}0$] direction. These spectroscopic features are consistent with a tetragonal-to-orthorhombic distortion below $T_2$=33K with the Mn$^{2+}$-O$^{2-}$ bond length expanded along the easy-axis [110] direction and contracted along the hard-axis [$1\overline{1}0$] direction (see illustrations, Fig. 6(b)).~\cite{ref32} More specifically, the greater intensity associated with the higher energy split mode in Fig. 6(b) shows that the contracted Mn$^{2+}$-O$^{2-}$ bond is oriented in the direction of the incident light polarization, i.e., along the [$1\overline{1}0$] direction; on the other hand, the weaker intensity of the lower energy split mode shows that the expanded Mn$^{2+}$-O$^{2-}$ bond is perpendicular to the incident light polarization, i.e., along the [110] direction, which is along the direction of the magnetic easy axis. This interpretation is confirmed by the \emph{T}=7K Raman spectrum obtained with light polarized along the [110] direction, shown in the inset of Fig. 6(a): in this scattering geometry, the 290 cm$^{-1}$ split mode has the higher intensity while the 300 cm$^{-1}$ split mode has the lower intensity, confirming that the expanded Mn$^{2+}$-O$^{2-}$ bond is oriented in the light polarization ([110]) direction in this scattering geometry.  These results demonstrate that the Raman spectrum can be used to identify not only the presence of a tetragonal-to-orthorhombic distortion at $T_2$=33 K in Mn$_3$O$_4$, but also the crystallographic orientation of the tetragonal distortion and magnetic easy axis.~\cite{ref32}

\begin{figure}[tb]
 \centerline{\includegraphics[width=8.5cm]{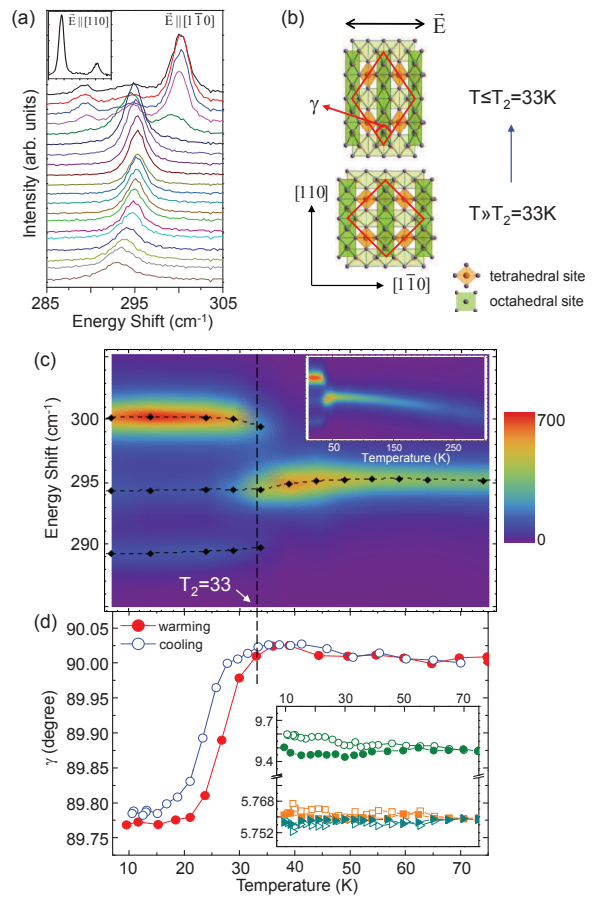}}
 \caption{\label{fig6}(a) Temperature dependence of the Raman spectra with increasing temperature from 7K to room temperature (top to bottom) for light polarized along the [$1\overline{1}0$] crystallographic direction. (inset) Raman spectrum at T=7K for light polarized along the [110] crystallographic direction. (b) Illustrations of Mn$_3$O$_4$ in the orthorhombic (T$<$$T_2$=33K) and tetragonal (T$>$$T_2$=33K) structures. (c) Contour plot of the $T_{2g}$ phonon mode intensity as functions of energy and increasing temperature. (inset) Contour plot of the $T_{2g}$ phonon mode intensity between 7-290K. (d) Temperature dependence of $\gamma$-the angle between the \emph{a}- and \emph{b}-axis directions-as functions of increasing (closed symbols) and decreasing (open symbols) temperature. (inset) Temperature dependence [in K] of lattice constants \emph{a} (squares), \emph{b} (triangles), and \emph{c} (circles) [in {\AA}] for Mn$_3$O$_4$ with decreasing (open symbols) and decreasing (closed symbols) temperature.}
\end{figure}

To provide further confirmation of the tetragonal-to-orthorhombic phase transition below $T_2$ in Mn$_3$O$_4$, temperature-dependent x-ray diffraction measurements were performed on the same Mn$_3$O$_4$ crystal: in particular, we measured the temperature dependence of the $\theta$, 2$\theta$, $\phi$ and $\chi$ angles of the (440) and (231) Bragg reflections. From these data, we deduced the temperature dependences of the lattice parameters $a$, $b$, $c$---where $a$ is taken along [100], $b$ is taken along [010] and $c$ is taken along [001]---as well as $\gamma$---the angle between $a$ and $b$---as functions of both increasing (``warming'') and decreasing (``cooling'') temperature; these parameters are summarized in Fig. 6(d).  While $\gamma$ exhibits an abrupt decrease near $T_2$=33K---indicating an abrupt decrease in the angle between \emph{a}- and \emph{b}-axis directions at this temperature, the lattice parameters \emph{a} and \emph{b} exhibit no temperature dependence, and the lattice parameter \emph{c} exhibits only a weak temperature dependence and hysteretic behavior. This behavior confirms that Mn$_3$O$_4$ exhibits a first-order tetragonal-to-orthorhombic structural phase transition near $T_2$=33K, as illustrated schematically in Fig. 6(b): this detailed information about the lattice parameters \emph{a}, \emph{b}, \emph{c}, and $\gamma$ enables us to identify---using a web-based program~\cite{ref34} with the atomic positions in the tetragonal structure~\cite{ref35}---the structure of Mn$_3$O$_4$ below $T_2$=33K as the orthorhombic structure with space group \emph{Fddd} (No. 70), where the principal axes for a unit cell are taken along [110], [$1\overline{1}0$], and [001].~\cite{ref32_1}

Our evidence for a structural distortion to a orthorhombic phase below $T_2$ is consistent with recent observations of a change in the bulk strain  $\Delta$$L$/$L$ below $T_2$ that is anisotropic in the \emph{ab}-plane.~\cite{ref24} Further, the orthorhombic distortion is also consistent with the spin structure previously reported for Mn$_3$O$_4$ below $T_2$: the coplanar spin structure of the doubled unit cell has spins lying within the ($1\overline{1}0$) plane with a net spin along [110], in which the Mn$^{3+}$ spins are canted by an angle ${\pm}{\Theta}_{YK}$ from the [$\overline{1}\overline{1}0$] direction. This spin canting is associated with a tilting of the $z^2$ orbitals of Mn$^{3+}$ toward the [$\overline{1}\overline{1}0$] direction due to spin-orbital coupling, which results in a tilting of Mn$^{3+}$ octahedra and an expansion of the Mn$^{2+}$-O$^{2-}$ bond length along the [110] direction.~\cite{ref21,ref24} We also note also that the absence of a structural distortion in the incommensurate magnetic phase regime between $T_2$=33K and $T_1$=39K (see Fig.6) is more consistent with the proposed spiral spin structure~\cite{ref21}---in which the Mn$^{3+}$ spins cant away from the ($1\overline{1}0$) plane---than with a sinusoidal spin structure:~\cite{ref22} the spiral spin structure is associated with an equivalence between the [110] and [$1\overline{1}0$] directions---which should favor a tetragonal structure---while the incommensurate sinusoidal spin structure is preferentially aligned within the ($1\overline{1}0$) plane, which should favor an orthorhombic structure.

Finally, the tetragonal-to-orthorhombic structural distortion at $T_2$ in Mn$_3$O$_4$ provides a mechanism for relieving the geometric frustration and allowing magnetic ordering in this spinel system, i.e., by defining a magnetic easy axis direction in the lower symmetry orthorhombic structure. This result implies that strong spin-lattice coupling is very important in determining the low-temperature phase behavior of spinels such as Mn$_3$O$_4$.

The tetragonal-to-orthorhombic structural change below $T_2$=33K, which is strongly coupled to the cell-doubled commensurate transition of Mn$_3$O$_4$, suggests that the magneto-structural phases in Mn$_3$O$_4$ can be manipulated by varying the magnitude and/or direction of an applied magnetic field. As described in the following sections, the distinctive Raman spectroscopic signatures associated with the different magneto-structural phases in Mn$_3$O$_4$---illustrated in Fig. 6(b)---offer a convenient method for investigating the diverse magnetostructural phases that can be induced in Mn$_3$O$_4$ with applied magnetic field and pressure.

\subsection{Magnetic field dependence}
The application of a magnetic field along different crystallographic directions in the ferrimagnetic phase of Mn$_3$O$_4$ provides a means of either enhancing the magnetic ordering tendencies of Mn$_3$O$_4$-for fields applied along the easy-axis [110] direction---or frustrating those ordering tendencies---for fields applied \emph{transverse} to the [110] direction.

Fig. 7 illustrates the field-induced structural phases of Mn$_3$O$_4$ for fields applied along the easy-axis [110] direction. Fig. 7(a) shows that the 295cm$^{-1}$ mode associated with the undistorted tetragonal phase (structure II in Fig. 7(b)) exhibits a field-induced splitting at T=39K$>$$T_2$ similar to that induced upon cooling below $T_2$=33 K in zero magnetic field (Fig. 6(a)).~\cite{ref32}  Figs. 7(c)-(f) show that this field-induced splitting is observed in the incommensurate magnetic phase regime $T_2$=33K$<$T$\leq$$T_1$=39K, which demonstrates that an applied magnetic field along the easy axis [110] direction increases the temperature at which the structural transition to the orthorhombic phase occurs in Mn$_3$O$_4$. However, for temperatures above \emph{T}=43K, no significant change in the spectrum is observed with applied field. These results show that, in the incommensurate magnetic phase, an applied magnetic field along the easy-axis [110] direction induces a tetragonal-to-orthorhombic distortion in Mn$_3$O$_4$ by forcing the Mn spins to order within the (110) plane and inducing the cell-doubled coplanar magnetic structure associated with the orthorhombic structure. This strong magnetoelastic response likely arises from a field-induced increase---via spin-orbit coupling---in the hybridization between the $d_{3z^2-r^2}$ and $d_{xy}$ orbitals of (octahedral) Mn$^{3+}$ for \textbf{H}$\parallel$[110]. Fig. 7(g) summarizes the different magnetostructural phases of Mn$_3$O$_4$ as functions of magnetic field and temperature for \textbf{H}$\parallel$[110].

\begin{figure}[tb]
 \centerline{\includegraphics[width=8.5cm]{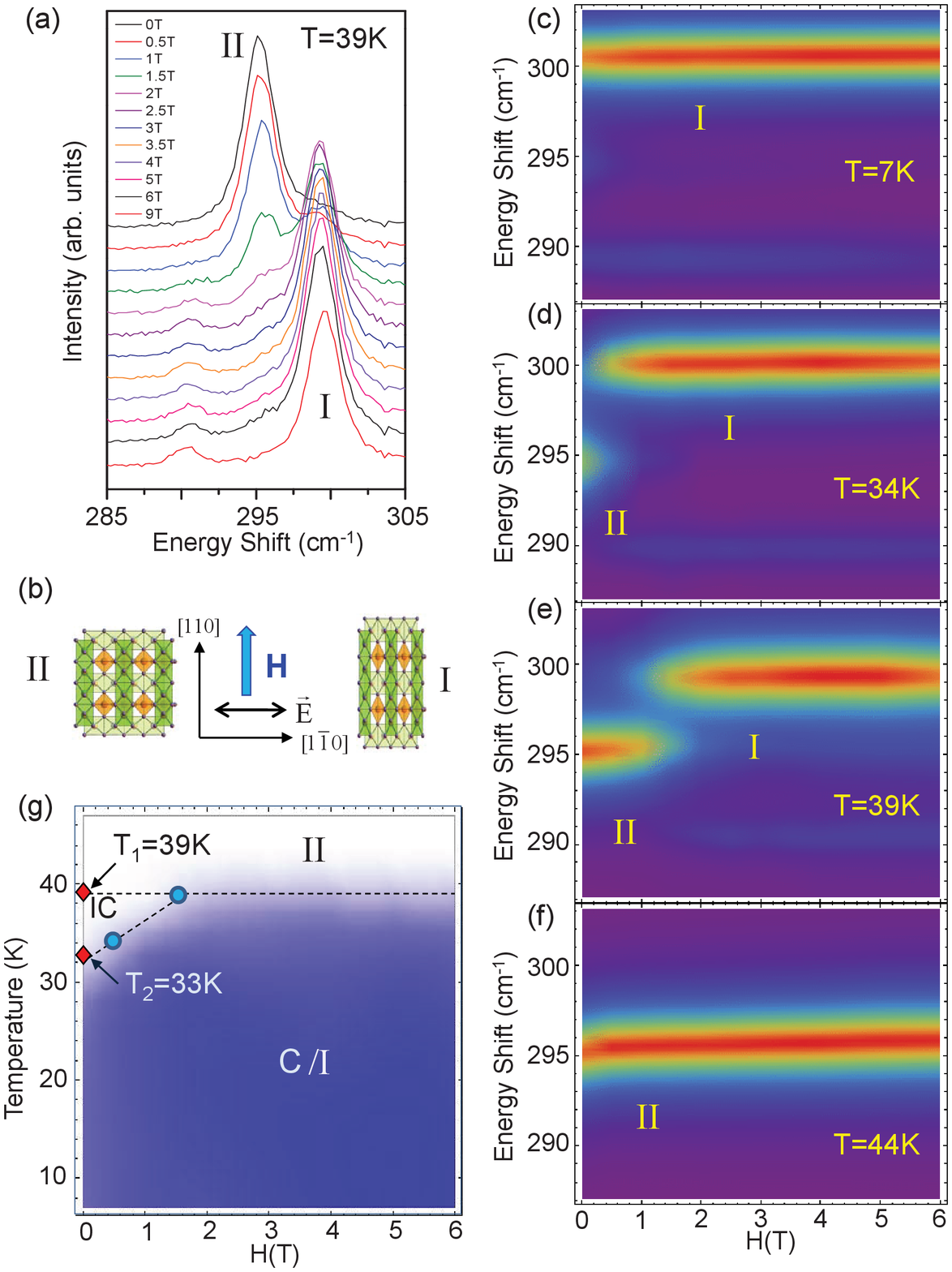}}
 \caption{\label{fig7}(a) Field dependence of the Raman spectra at T=39K for \textbf{H}$\parallel$[110]. (b) Illustrations of the (II) low-field tetragonal and (I) high-field orthorhombic structures of Mn$_3$O$_4$. (c)-(f) Contour plots of the $T_{2g}$ phonon mode intensities as functions of energy and field for \textbf{H}$\parallel$[110] at (c) \emph{T}=7K, (d) 34K, (e) 39K, and (f) 44K. (g) \emph{H-T} phase diagram for \textbf{H}$\parallel$[110]); blue=structure I, white=structure II, IC=incommensurate magnetic phase, C=commensurate (cell-doubled) magnetic phase.}
\end{figure}

A richer magneto-structural phase diagram is apparent for a magnetic field applied in a direction \emph{transverse} to the easy-axis direction, \textbf{H}$\parallel$[$1\overline{1}0$], i.e., in a direction that places the applied field in competition with the ferrimagnetic ordering direction in Mn$_3$O$_4$. Figs. 8(a) and (c) display the field dependence of the phonon spectrum at \emph{T}=7K ($\ll$ $T_2$ = 33K) with increasing fields for \textbf{H}$\parallel$[$1\overline{1}0$], showing that the relative scattering intensities of the 290 cm$^{-1}$ and 300 cm$^{-1}$ modes---i.e., the $T_{2g}$(1) modes that split upon cooling below $T_2$=33 K at \emph{H}=0 T---are reversed by applying a transverse field with a sufficiently high magnetic field strength (\emph{H} $>$ 5.5 T), reflecting a 90$^{\circ}$ rotation of the long axis of the orthorhombic distortion (and the magnetic easy axis) from the [110] to the [$1\overline{1}0$] directions (i.e., from structure I to structure III in Fig. 8(b)). Notably, there is significant hysteresis associated with this field-induced distortion, indicating that these field-induced phase transitions are first order: Fig. 8(d) shows that, with deceasing fields from \emph{H}=6 T to 0 T, the high-field reoriented phase (\textbf{M}$\parallel$[$1\overline{1}0$]) does not relax back to the initial low-field phase (\textbf{M}$\parallel$[110]) until the field is reduced to $\sim$0.5 T.

\begin{figure}[tb]
 \centerline{\includegraphics[width=8.5cm]{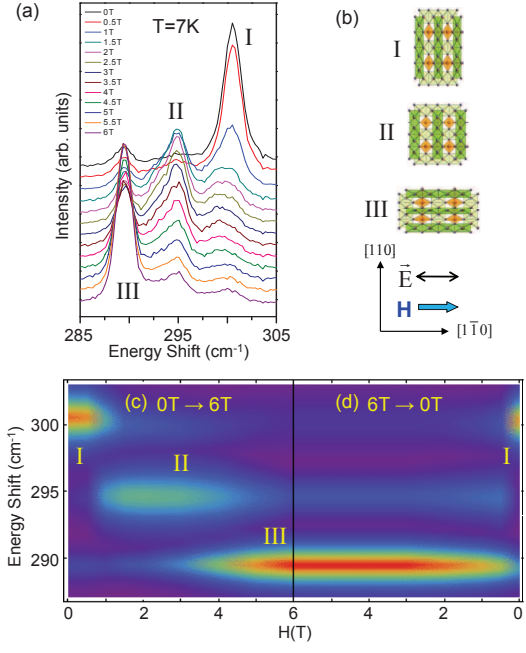}}
 \caption{\label{fig8}(a) Field dependence of the Raman spectra at \emph{T}=7K for \textbf{H}$\parallel$[$1\overline{1}0$]. (b) Illustrations of the Mn$_3$O$_4$ structure in the (top) the low-field orthorhombic, (middle) intermediate-field tetragonal, and (bottom) high-field orthorhombic phases Contour plots of the $T_{2g}$ phonon mode intensities as functions of energy and field at (a) \emph{T}=7K, with increasing fields and decreasing field.}
\end{figure}

Most remarkably, however, Figs. 8(a) and (c) shows that the field-induced transition between low- and high-field phases---i.e., between structures I and III in Fig. 8(b)---occurs via an intermediate field regime (1T $<$ \emph{H} $<$ 4T) in which the phonon splitting disappears and the $\sim$295cm$^{-1}$ mode associated with the undistorted tetragonal phase is dominant. This spectroscopic signature indicates that Mn$_3$O$_4$ adopts a more symmetric (tetragonal) structural configuration at intermediate values of the applied magnetic field for T$<$$T_2$=33K, in order to resolve the frustration imposed on the Mn spins by the  competition between the internal (along [110]) and transverse external (along [$1\overline{1}0$]) magnetic fields in Mn$_3$O$_4$. This field-induced symmetric structural configuration is expected to be associated with a disordered spin configuration, by supporting an equivalency between [110] and [$1\overline{1}0$] Mn spin orientations that is similar to that observed in the high temperature paramagnetic phase. Indeed, magnetization measurements of Mn$_3$O$_4$ by Dwight and Menyuk revealed that the application of a $\sim$1 T field transverse to the easy axis direction of Mn$_3$O$_4$ at 4.2 K resulted in slow spin relaxation processes consistent with a disordered spin system.~\cite{ref36} Fig. 9(a)-(d) show the field dependence at different temperatures, demonstrating that, with increasing temperatures, this field-induced tetragonal regime becomes increasingly dominant with increasing applied field along the [$1\overline{1}0$] direction. The full \emph{H-T} phase diagram for Mn$_3$O$_4$ with \textbf{H}$\parallel$[$1\overline{1}0$] inferred from our Raman results is summarized in Fig. 9(f).

\begin{figure}[tb]
 \centerline{\includegraphics[width=8.5cm]{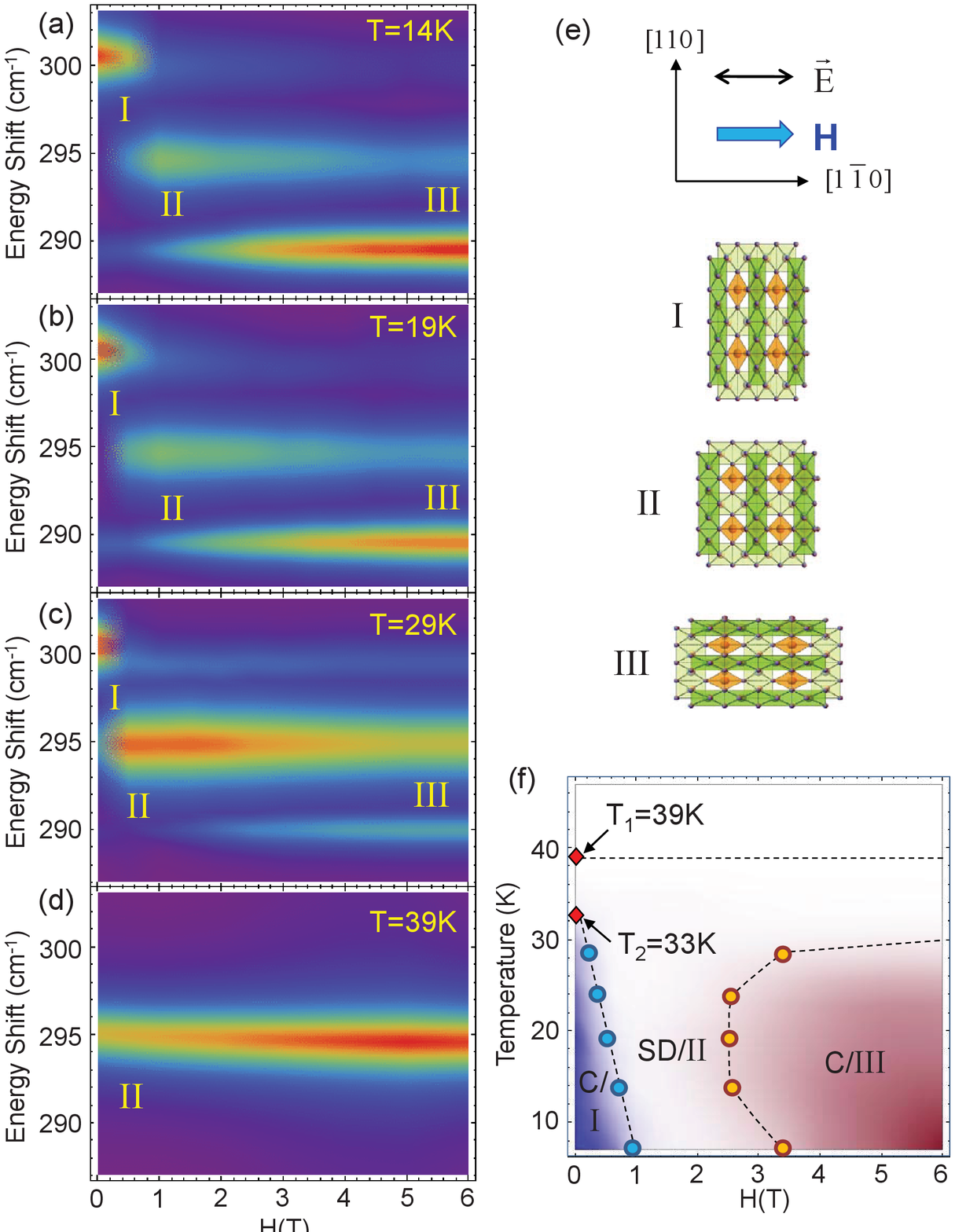}}
 \caption{\label{fig9}Contour plots of the $T_{2g}$ phonon mode intensities as functions of energy and field for \textbf{H}$\parallel$[$1\overline{1}0$] (a) \emph{T}=14K, (b) 19K, (c) 29K, and (d) 39K. (e) Illustrations of the Mn$_3$O$_4$ structure in (I) the orthorhombic with magnetic easy axis along [110], (II) tetragonal, and (III) orthorhombic structure of Mn$_3$O$_4$ with magnetic easy axis along [$1\overline{1}0$]. (f) \emph{H-T} phase diagram for \textbf{H}$\parallel$[$1\overline{1}0$]; blue=structure I, white=structure II, red=structure III, SD=spin-disordered phase, and C=commensurate (cell-doubled) magnetic phase.}
\end{figure}

This three-stage field dependence of the magnetostructural phases in Mn$_3$O$_4$ for \textbf{H}$\parallel$[$1\overline{1}0$] is also confirmed by field-dependent x-ray scattering measurements of the $\theta$, 2$\theta$, and $\phi$ angles associated with the (220) and (440) Bragg reflections. Figure 10 shows the intensities of the (440) reflection peak as functions of magnetic field and 2$\theta$ angle for different temperatures. As seen in Figure 10, in the low-field regime (\emph{H}$<$1T) for  \emph{T} $<$ $T_2$ (=33K), the 2$\theta$ angle has a value consistent with an expanded MnO bond length along the [110] direction; on the other hand, in the high-field regime (\emph{H}$>$3T), the 2$\theta$ angle exhibits a larger value, consistent with a contracted MnO bond length along the [110] directions; this transition  occurs  via an intermediate field regime in which the 2$\theta$ angle has an intermediate value consistent with an undistorted lattice. Within each regime, the 2$\theta$ angle doesn't show a significant change with changing magnetic field, indicating that field dependent changes in the Mn$_3$O$_4$ are not continuous, but occur via discontinuous jumps between the three distinct structures illustrated in Fig. 10. From these field-dependent x-ray results at different temperatures, we're able to map out a magnetostructural phase diagram of Mn$_3$O$_4$ for \textbf{H}$\parallel$[$1\overline{1}0$] that is consistent with the Raman results summarized in Fig. 9.

\begin{figure}[tb]
 \centerline{\includegraphics[width=8.5cm]{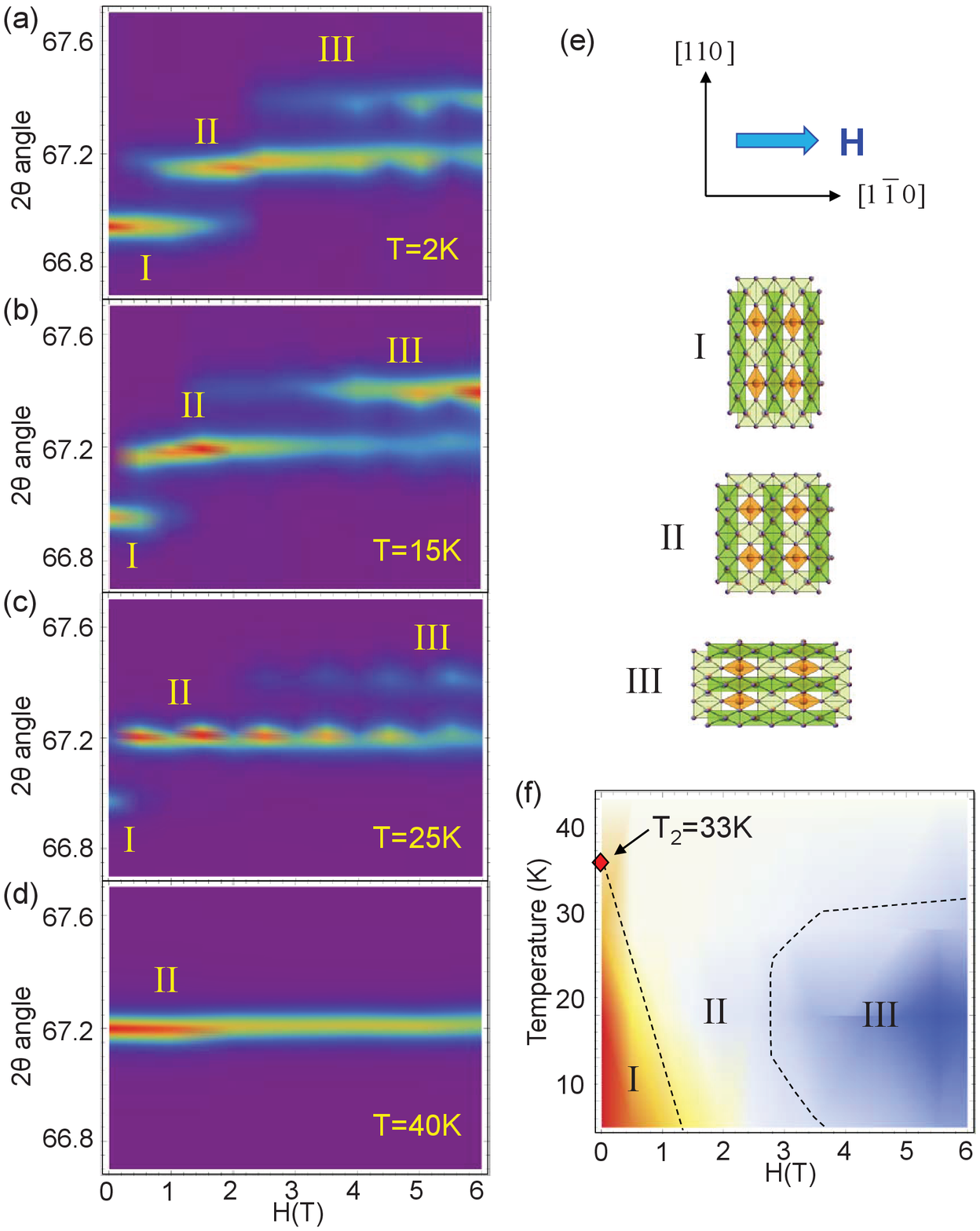}}
 \caption{\label{fig10}Contour plots of the (440) reflection intensities as functions of 2$\theta$ angle and field for \textbf{H}$\parallel$[$1\overline{1}0$]. (a) \emph{T}=2K, (b) 15K, (c) 25K, and (d) 40K. (e) Illustrations of the Mn$_3$O$_4$ structure in (I) the orthorhombic with magnetic easy axis along [110], (II) tetragonal, and (III) orthorhombic structure of Mn$_3$O$_4$ with magnetic easy axis along [$1\overline{1}0$]. (f) \emph{H-T} phase diagram for \textbf{H}$\parallel$[$1\overline{1}0$] based on the x-ray measurement results, and illustrations for the corresponding structures.}
\end{figure}

These magnetic-field-dependent Raman and x-ray results illustrate that the magnetic frustration and high structural symmetry that is removed via a low-temperature structural transition in Mn$_3$O$_4$ can be recovered at arbitrarily low temperatures by applying a transverse magnetic field that competes with the internal magnetic field. This is indicative of a discontinuous quantum (\emph{T}$\sim$0K) phase transition from a structurally distorted, magnetically ordered state to an undistorted, spin-disordered state. This transition reveals a novel mechanism---made possible by strong spin-orbital coupling---by which a transverse applied field in the ferrimagnetic state can induce a frustrated spin/orbital state at \emph{T}$\sim$0 by first inducing a more symmetric \emph{structural} configuration that is compatible with degenerate spin and orbital states. This route to frustrating spin- and orbital-order at \emph{T}$\sim$0 involves a balancing of elastic and magnetic energies that has not been generally considered when modeling field-induced spin/orbital frustration,~\cite{ref20} and suggests that a transverse applied field can negate the effects of spin-lattice coupling that generates the low temperature distortion for \emph{H}=0 T in Mn$_3$O$_4$. Interestingly, our results suggest that the transverse applied field plays a similar role in generating orbital-spin frustration in Mn$_3$O$_4$ that hydrostatic pressure has been shown to play in other orbital-ordered systems, i.e., by frustrating orbital order through the creation of a more symmetric structural configuration that supports degenerate orbital-spin states.~\cite{ref37}

These results also provide strong evidence that the magnetic-field-tunable strain and magnetodielectric behavior observed previously in Mn$_3$O$_4$~\cite{ref23,ref24} is microscopically associated with magnetic-field-controlled orthorhombic-to-tetragonal distortions, which likely result from field-induced mixing of the hybridization between the $d_{3z^2-r^2}$ and $d_{xy}$ orbitals of octahedral Mn$^{3+}$. These magnetic-field-induced structural transitions cause phonon frequency changes that modify the dielectric response according to the Lyddane-Sachs-Teller (LST) relationship,~\cite{ref38} $\varepsilon$(0)/$\varepsilon$($\infty$)=${\omega}^2_L$/${\omega}^2_T$, where $\varepsilon$(0) ($\varepsilon$($\infty$)) is the zero-frequency (high-frequency) dielectric response of the material, and ${\omega}^2_L$(${\omega}^2_T$) is the long-wavelength longitudinal (transverse) phonon frequency.

\begin{figure}[tb]
 \centerline{\includegraphics[width=8.5cm]{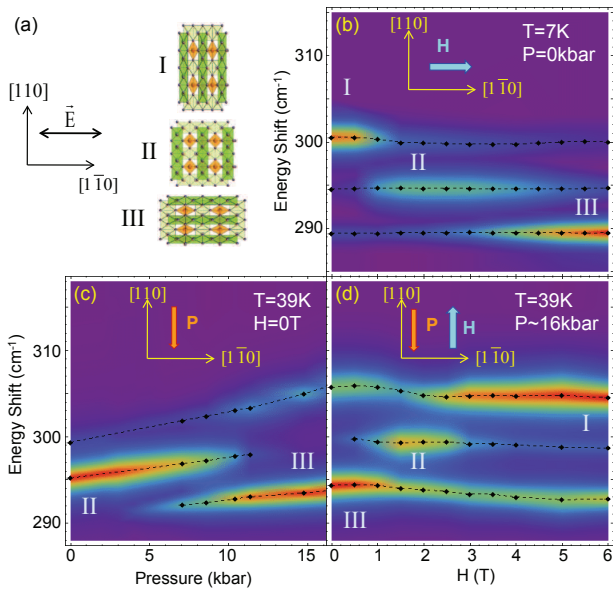}}
 \caption{\label{fig11}(a) Illustrations of (I) the orthorhombic structure of Mn$_3$O$_4$ with magnetic easy axis along [110], (II) the tetragonal structure of Mn$_3$O$_4$, and (III) the orthorhombic structure of Mn$_3$O$_4$ with magnetic easy axis along [$1\overline{1}0$]. (b) Contour plot of the $T_{2g}$ phonon mode intensity as functions of energy and applied magnetic field (\textbf{H}$\parallel$[$1\overline{1}0$]), where red=maximum counts and blue=0 counts. (c) Contour plot of the $T_{2g}$ phonon mode intensities as functions of energy and applied pressure (\textbf{P}$\parallel$[110]) for \emph{T}=39K and \emph{H}=0T. (d) Contour plot of the $T_{2g}$ phonon mode intensities as functions of energy and applied magnetic field (\textbf{H}$\parallel$[110]) for \emph{T}=39K and \emph{P}=16kbar.  Filled diamonds represent the $T_{2g}$ mode energies and the dashed lines are guides to the eye.}
\end{figure}

\subsection{Pressure dependence}
The strong sensitivity of the structural phases of Mn$_3$O$_4$ to applied magnetic field also suggests that the magneto-structural and -dielectric properties of this material can be controlled using applied pressure.  Fig. 11(c) shows the pressure dependence of the $T_{2g}$ mode at \emph{T}=39 K and \emph{H}=0 T.  At low pressures (\emph{P}$<$7 kbar), the $\sim$295 cm$^{-1}$ degenerate $T_{2g}$ phonon mode---associated with the undistorted tetragonal phase (structure II in Fig. 11(a))---shifts to higher frequencies with applied pressure (with a rate $d{\omega}_{\circ}$/$dP$$\sim$0.25 cm$^{-1}$/kbar, where ${\omega}_{\circ}$ is the $T_{2g}$ phonon frequency), indicating a simple pressure-induced reduction of the unit cell volume with no structural transition.  However, above \emph{P}*$\sim$8.5kbar, the $T_{2g}$ mode exhibits a pressure-induced splitting into an intense lower frequency mode and a weak higher frequency mode: This is the spectroscopic signature of a pressure-induced transition at \emph{T}=39 K from the tetragonal phase (structure II in Fig. 11(a)) to an orthorhombic phase with the long axis of the distortion (and the magnetic easy axis) \emph{parallel} to the polarization direction of the incident electric field, \textbf{E}$_i$$\parallel$[$1\overline{1}0$] (structure III in Fig. 11(a)).  The results in Fig. 11(c) demonstrate that the magnetic easy axis (and the long axis of the orthorhombic distortion) of Mn$_3$O$_4$ is rotated from the [110] direction to the [$1\overline{1}0$] direction with sufficiently high pressures applied in the [110] direction.  This pressure-induced rotation of the magnetic easy axis is confirmed in Fig. 11(d), which shows that a magnetic field applied \emph{transverse} to the [$1\overline{1}0$] direction---i.e., transverse to the pressure-induced magnetic easy axis direction (i.e., \textbf{H} $\perp$\textbf{M}$\parallel$[$1\overline{1}0$])---\emph{reverses} the magnetostructural changes induced with pressure.  Specifically, Fig. 11(d) shows that an applied field \textbf{H}$\parallel$[110] (with \emph{P}=16 kbar and \emph{T}=39 K) induces a transition at $H_{c_1}$$\sim$1.5 T from an orthorhombic phase with magnetic easy axis parallel to [$1\overline{1}0$] (structure III in Fig. 11(a)) to a tetragonal phase (structure II in Fig. 11(a)), as evidenced by the field-induced collapse of the two split phonon modes to a single mode near 295 cm$^{-1}$.  At still higher applied fields (\textbf{H}$\parallel$[110] with \emph{P}=16 kbar and \emph{T}=39 K), there is a second transition at $H_{c_2}$$\sim$3 T between a tetragonal phase (structure II in Fig. 11(a)) and an orthorhombic phase with magnetic easy axis parallel to [110] (structure I in Fig. 11(a)); this structural change is identified by the splitting of the $\sim$295 cm$^{-1}$ phonon mode into an intense higher frequency mode and a weak lower frequency mode.

Figs. 12(a)-(c) display the pressure-dependences of the $\sim$295 cm$^{-1}$ $T_{2g}$ phonon mode at different temperatures in Mn$_3$O$_4$, showing that applied pressure induces a tetragonal-to-orthorhombic transition---and orients the magnetic easy axis along the [$1\overline{1}0$] direction---at successively higher critical pressures with increasing temperature, i.e., from \emph{P}*$\sim$8.5kbar at 39 K to \emph{P}*$\sim$27 kbar at 54 K.  Fig. 12(e) summarizes the \emph{PT} phase diagram of Mn$_3$O$_4$ inferred from these results:  Below \emph{T}=33 K and at pressures \emph{P}$<$5 kbar, Mn$_3$O$_4$ has an orthorhombic phase with a magnetic easy axis along the [110] direction (structure I in Fig. 12(d)).  However, above a critical pressure \emph{P}*---whose value increases with temperature with a rate $\Delta$$P$/$\Delta$$T$$\sim$1.2 kbar/K (orange circles in Fig. 12(e)) above \emph{T}=33K---there is a transition to an orthorhombic phase with magnetic easy axis along the [$1\overline{1}0$] direction (structure III in Fig. 12(d)). Between these two phases is a tetragonal phase (structure II in Fig. 12(d)) that does not exhibit commensurate long-range magnetic order.

\begin{figure}[tb]
 \centerline{\includegraphics[width=8.5cm]{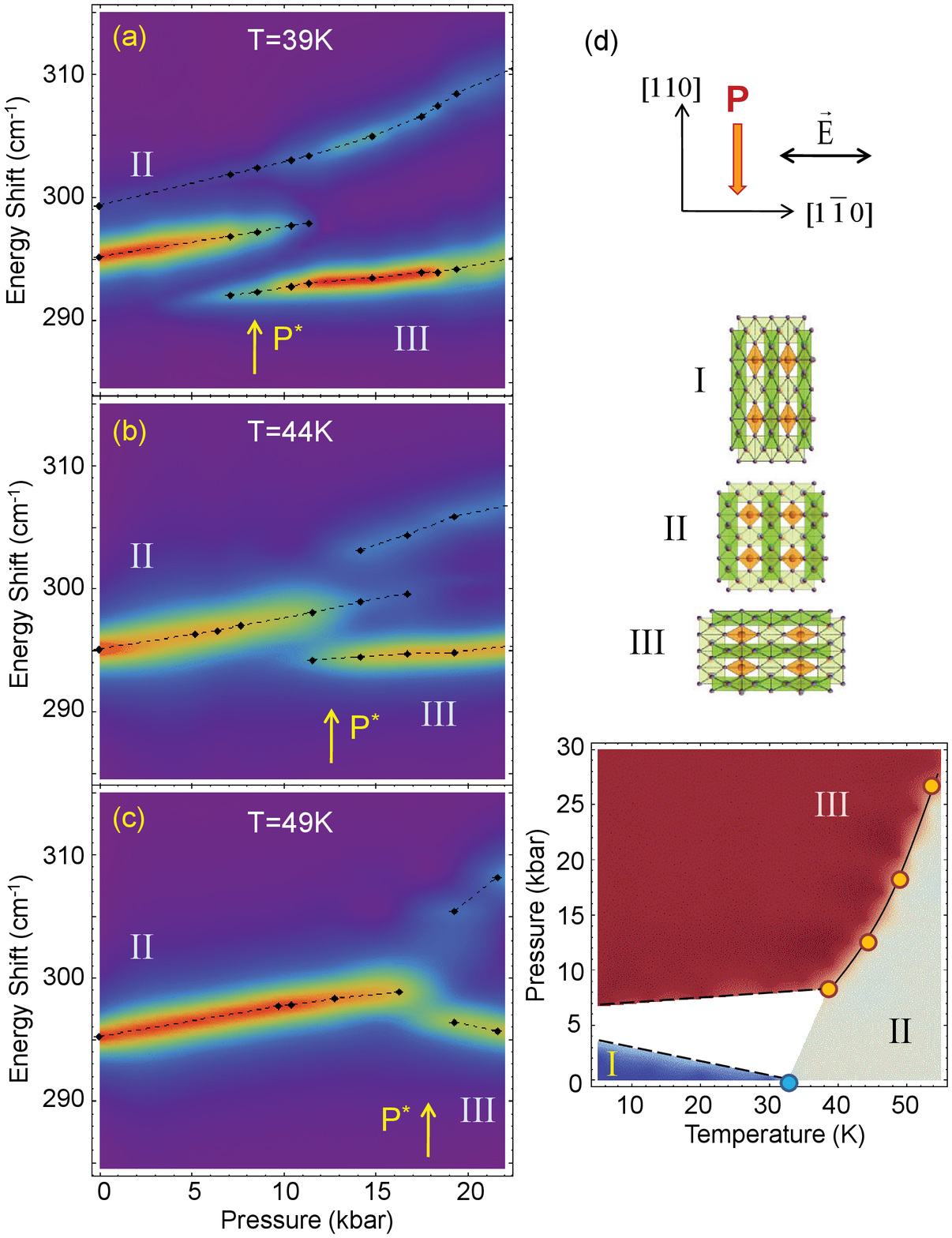}}
 \caption{\label{fig12}Contour plots of the $T_{2g}$ phonon mode intensities as functions of energy and applied pressure (\textbf{P}$\parallel$[110]) for \emph{H}=0T and (a) \emph{T}=39K, (b) \emph{T}=44K, and (c) \emph{T}=49K, where red=maximum counts and blue=0 counts.  Filled diamonds represent the $T_{2g}$ mode energies and the dashed lines are guides to the eye.  (d) Illustrations of (I) the orthorhombic structure of Mn$_3$O$_4$ with magnetic easy axis along [110], (II) the tetragonal structure of Mn$_3$O$_4$, and (III) the orthorhombic structure of Mn$_3$O$_4$ with magnetic easy axis along [$1\overline{1}0$]. (e) \emph{PT}-phase diagram inferred from results in (a)-(c).  The blue circle represents the measured critical temperature between magnetostructural phase regions I (blue) and II (gray), while the orange circles represent measured critical pressures between phase regions III (red) and II (gray).  Dashed lines are estimated phase boundary lines, and the white area is a region in which the detailed phase behavior is unknown.}
\end{figure}

The pressure-dependent results in Figs. 11 and 12 confirm that applied pressure influences the magnetic easy axis direction of Mn$_3$O$_4$---and increases the temperature at which magneto-structural phase changes and magneto-dielectric behavior are observed---by tuning the magnetocrystalline anisotropy of the material; further, pressure orients the magnetic easy axis in the direction of a strain field that is transverse to the direction of the applied pressure. One interesting implication of the results summarized in Fig. 12(e) is that significant piezo-dielectric effects---associated with pressure-induced magnetostructural changes in Mn$_3$O$_4$---should be observed near the phase boundaries shown in Fig. 12(e).  These results also suggest that it should be possible to engineer the magneto-structural phases shown in Fig. 12(e)---and thereby control the magneto-dielectric properties that accompany these magnetostructural phase changes---via compressive epitaxial strain in oriented films of Mn$_3$O$_4$. Indeed, a previous study by Gorbenko et al.~\cite{ref39} reported that  Mn$_3$O$_4$ films grown on MgO substrates exhibit a cubic structure rather than a tetragonal structure expected of bulk Mn$_3$O$_4$ samples below \emph{T}$\sim$1440K, presumably because epitaxial strain suppresses the Jahn-Teller (JT) distortions responsible for the cubic-to-tetragonal transition.~\cite{ref39} Our high pressure results in Figs. 11 and 12 further suggest that other magnetostructural phases, e.g., orthorhombic distortions with the long axis along a certain preferred directions, might be attainable in Mn$_3$O$_4$ films by using appropriate substrates.

\begin{figure}[tb]
 \centerline{\includegraphics[width=8.5cm]{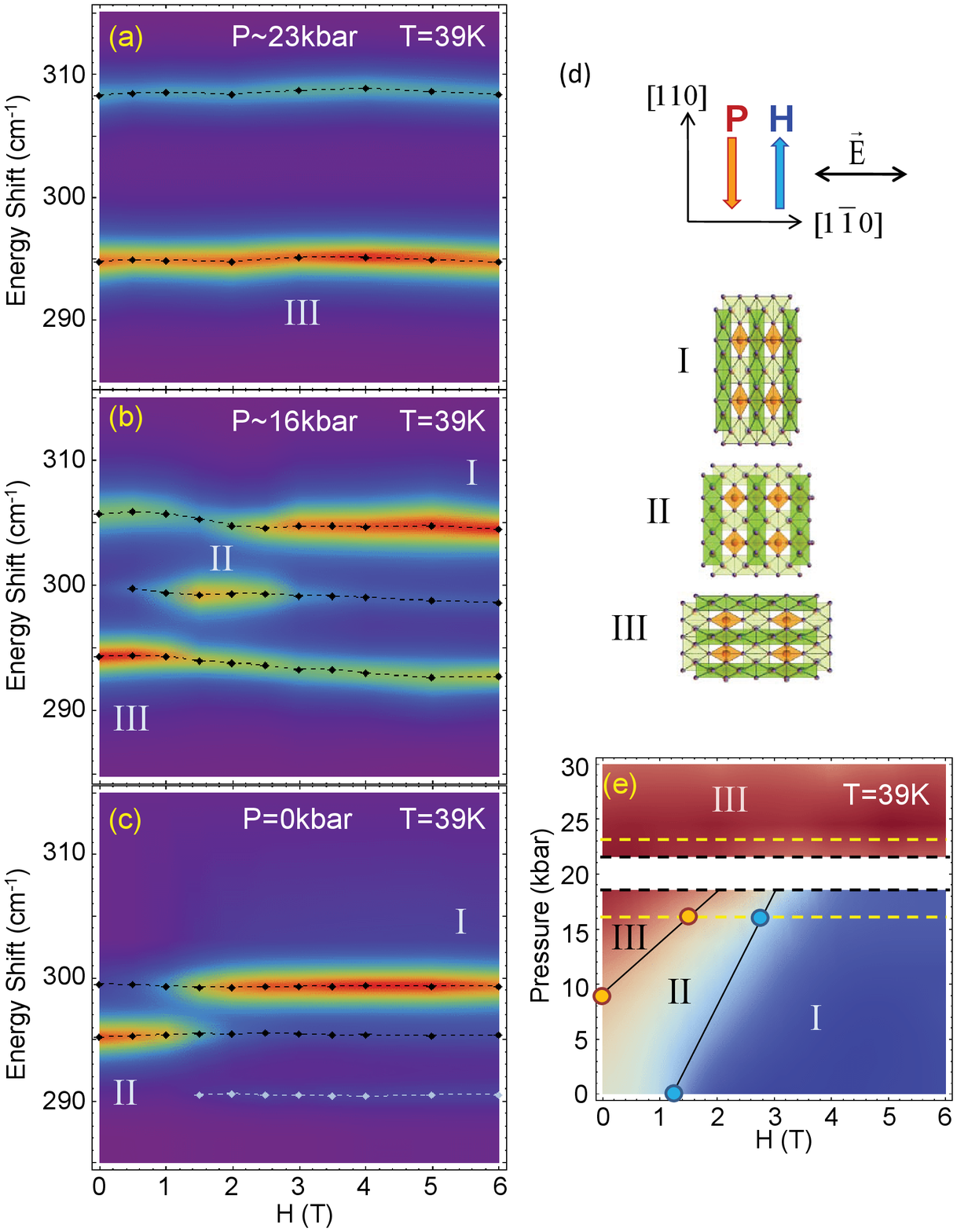}}
 \caption{\label{fig13}Contour plots of the $T_{2g}$ phonon mode intensities as functions of energy and applied field (\textbf{H}$\parallel$[110]) for \emph{T}=39K and (a) \emph{P}=23kbar, (b) \emph{P}=16kbar, and (c) \emph{P}=0kbar, where red=maximum counts and blue=0 counts.  Filled diamonds represent the $T_{2g}$ mode energies and the dashed lines are guides to the eye.  (d) Illustrations of (I) the orthorhombic structure of Mn$_3$O$_4$ with magnetic easy axis along [110], (II) the tetragonal structure of Mn$_3$O$_4$, and (III) the orthorhombic structure of Mn$_3$O$_4$ with magnetic easy axis along [$1\overline{1}0$]. (e) \emph{PH}-phase diagram inferred from results in (a)-(c).  Orange circles represent measured critical pressures between magnetostructural phase regions III (red) and II (gray), while blue circles represent measured critical pressures between magnetostructural phase regions II (gray) and I (blue).  Dashed yellow lines indicate pressures at which magnetic field sweeps were obtained, and the white area is a region in which the detailed phase behavior is unknown.}
\end{figure}

To explore the rich magnetostructural phases that can be obtained in Mn$_3$O$_4$ as functions of both magnetic field and pressure, Figs. 13(a)-(c) show the magnetic-field-dependence of the structural phases of Mn$_3$O$_4$ at different values of applied pressure for \textbf{H}$\parallel$[110]. Fig. 13(e) summarizes the \emph{PH} phase diagram of Mn$_3$O$_4$ at \emph{T}=39 K inferred from these results:  for low values of the pressure, \emph{P}$>$8.5kbar, there is a field-induced transition from the tetragonal phase (structure II in Fig. 13(d)) to an orthorhombic phase with magnetic easy axis along the [110] direction. However, for \emph{P}$>$8.5kbar, increasing applied field (\textbf{H}$\parallel$[110]) induces magnetostructural transitions, first, between an orthorhombic structure with magnetic easy axis along the [$1\overline{1}0$] direction (structure III in Fig. 13(d)) and a tetragonal structure (structure II in Fig. 13(d)), and second, between a magnetically frustrated tetragonal structure and an orthorhombic structure with magnetic easy axis along the [110] direction (structure I in Fig. 13(d)).  Fig. 13(e) shows that the magnetically frustrated tetragonal phase regime (structure II) is pushed to higher fields---and has a somewhat narrower field range---with increasing pressure. For \emph{P}$\geq$23 kbar, no magnetostructural transition is observed within our observed field range, suggesting that orthorhombic structure III becomes clamped by the strong strain field---and is therefore unresponsive to an applied field---in this pressure and field range.

Several noteworthy features of the \emph{PH} phase diagram are apparent in Fig. 13(e): the largest magnetoelastic and magnetodielectric changes in Mn$_3$O$_4$ are expected at the phase boundaries shown in Fig. 13(e), where the most rapid changes in structure and phonon frequencies are observed with field. Thus, Fig. 13(e) reveals both the pressure range over which magnetodielectric properties of Mn$_3$O$_4$ can be tuned and the pressure and field values at which the largest magnetodielectric changes are expected. Additionally, the phase transition boundaries in Fig. 13(e) are both first-order transitions, as evidenced by hysteresis in the field-dependent evolution of the phonon spectra when cycling the magnetic field. However, the phase diagram in Fig. 13(e) suggests that there is a path in the \emph{PH} phase diagram along which a direct transition between orthorhombic structures I and III might be induced, e.g., by first applying a magnetic field along the [110] direction to induce magnetostructural phase I, then deforming the crystal with pressure---and continuously rotating the magnetic easy axis---between the [110] and [$1\overline{1}0$] directions.  Efforts are currently ongoing to see whether a continuous quantum phase transition between magnetostructural phases can be induced in Mn$_3$O$_4$ with the simultaneous application of magnetic field and pressure.

\section{Summary and conclusions}
In summary, the magnetostructural phases of Mn$_3$O$_4$ were carefully examined as functions of temperature, pressure, and magnetic field by studying the Raman-active phonon spectrum, which affords detailed information about specific parts of the Mn$_3$O$_4$ structure. This study revealed several new details of the magnetostructural phases and properties of Mn$_3$O$_4$. First, temperature-dependent measurements show a distinct splitting of Raman-active phonons below $T_2$=33K, indicating that a tetragonal-to-orthorhombic structural distortion is associated with the cell-doubled commensurate magnetic ordering transition. This result provides a specific microscopic mechanism by which geometric frustration is resolved---and magnetic ordering is allowed---in Mn$_3$O$_4$.  Secondly, the field-induced phonon splitting and split-mode switching observed in the field-dependent measurements on Mn$_3$O$_4$ illustrates not only that the tetragonal-to-orthorhombic transition can be sensitively controlled with an applied field, but that the direction of the distortion can be tuned with an applied field.  Most significantly, we find evidence for a quantum (\emph{T}$\sim$0) phase transition to a structurally isotropic, disordered spin/orbital state for magnetic fields applied transverse to the ferrimagnetic easy-axis direction, i.e., for \textbf{H}$\parallel$[$1\overline{1}0$], which reflects a field-tuned degeneracy between competing magnetostructural configurations, and demonstrates the important role that magnetoelastic interactions can play in frustrating spin and orbital order at \emph{T}=0. Finally, our pressure-dependent Raman measurements show that the magnetic easy axis direction in Mn$_3$O$_4$ can be controlled---and the ferrimagnetic transition temperature increased---with applied pressure. Our combined pressure- and magnetic-field-tuned Raman measurements reveal a rich magnetostructural phase diagram---including a pressure- and field-induced magnetically frustrated tetragonal phase in the \emph{PH} phase diagram---that can be induced in Mn$_3$O$_4$ with a combination of applied pressure and field. Efforts are currently ongoing to see whether a continuous quantum phase transition between magnetostructural phases can be induced in Mn$_3$O$_4$ with the simultaneous application of magnetic field and pressure.

\begin{acknowledgments}
Research supported by the U.S. Department of Energy, Office of Basic Energy Sciences, Division of Materials Sciences and Engineering under Award DE-FG02-07ER46453.  Work by M. K. was supported by the National Science Foundation under Grant NSF DMR 08-56321.  Use of the National Synchrotron Light Source, Brookhaven National Laboratory, was supported by the U.S. Department of Energy, Office of Science, Office of Basic Energy Sciences, under Contract No. DE-AC02-98CH10886. We thank Gang Cao and Oleksandr Korneta of the U. of Kentucky for use of their Magnetic Properties Measurement System for the magnetic susceptibility measurements described here.
\end{acknowledgments}


\begin{thebibliography}{}
\bibitem{ref30}We follow Ref.~\onlinecite{ref29} and Ref.~\onlinecite{ref31} in using the cubic space group notation to index the Raman-active phonon modes in Mn$_3$O$_4$. This is justified by the fact that Ref.~\onlinecite{ref28} and Ref.~\onlinecite{ref29} showed that the Raman spectra of tetragonal spinels Fe$_3$O$_4$, Mn$_3$O$_4$, Li$_2$Mn$_2$O$_4$, and AMn$_2$O$_4$ (A=Mn, Zn, Mg) exhibited phonon spectra similar to spectra observed in cubic spinels.
\end{thebibliography}
\end{document}